\documentclass{article}

\usepackage{microtype}
\usepackage{graphicx}
\usepackage{subfigure}
\usepackage{booktabs} 
\usepackage{multirow}

\usepackage[acronym]{glossaries}
\makeglossaries

\newacronym{llm}{LLM}{large language model}
\newacronym{gpu}{GPU}{Graphics Processing Unit}
\newacronym{dsmem}{DSMEM}{distributed shared memory}
\newacronym{gpc}{GPC}{Graphics Processing Cluster}
\newacronym{tpc}{TPC}{Texture Processing Cluster}
\newacronym{sm}{SM}{Streaming Multiprocessor}
\newacronym{dvfs}{DVFS}{Dynamic Voltage and Frequency Scaling}
\newacronym{dnn}{DNN}{Deep Neural Network}
\newacronym{tma}{TMA}{Tensor Memory Accelerator}
\newacronym{isa}{ISA}{Instruction Set Architecture}
\newacronym{tdp}{TDP}{thermal design power}
\newacronym{ncu}{NCU}{NVIDIA Nsight Compute}
\newacronym{sku}{SKU}{stock-keeping unit}
\newacronym{noc}{NoC}{Network-on-Chip}
\newacronym{ic}{IC}{integrated circuit}

\usepackage{hyperref}

\usepackage[accepted]{mlsys2025}

\makeatletter
\renewcommand{\Notice@String}{}
\makeatother
\hypersetup{
    pdfsubject={Demystifying the Impact of Physical Non-Idealities on Kernel Performance}
}

\mlsystitlerunning{Demystifying the Impact of Physical Non-Idealities on Kernel Performance}

\begin{document}

\twocolumn[
\mlsystitle{Demystifying the Impact of Physical Non-Idealities on Kernel Performance}



\mlsyssetsymbol{equal}{*}

\begin{mlsysauthorlist}
\mlsysauthor{Supun Kuruppu}{nus}
\mlsysauthor{Pranav Dangi}{nus}
\mlsysauthor{Wei Li}{nus}
\mlsysauthor{Zhenyu Bai}{nus}
\mlsysauthor{Tulika Mitra}{nus}
\end{mlsysauthorlist}

\mlsysaffiliation{nus}{Department of Computer Science, National University of Singapore, Singapore}

\mlsyscorrespondingauthor{Zhenyu Bai}{zhenyu.bai@nus.edu.sg}

\mlsyskeywords{Kernel Performance Engineering, Manufacturing Defects, Thermal Design Power, Floorsweeping, Dynamic Voltage and Frequency Scaling, GPU, Accelerators, Distributed Shared Memory, Wave-Aware Optimizations}


\begin{abstract}

\glsunset{gpu}

Physical non-idealities in \gls{ic} design, fabrication, and operation are concealed by conventional hardware abstractions, causing unexpected performance variations in kernels.
For large-scale workloads, such as AI training and inference, performance unpredictability can lead to substantial losses.
We investigate this problem in NVIDIA H100 \glspl{gpu} with their \gls{dsmem} architectural feature that reduces data movement from the L2 cache to \glspl{sm} with multicast loads, and uncover that floorsweeping restricts the number of \glspl{sm} usable for clustered execution, increasing the number of execution waves of kernels.
To alleviate this performance penalty, we introduce the wave-aware clustering strategy that accounts for these resource constraints and improves the throughput of the Hopper version of FlashAttention 4 by up to 5.30\% over the unclustered baseline.
Furthermore, we discover that the power constraint imposed by the \gls{tdp} limit causes the \gls{dsmem}-based memory bandwidth optimization to improve the throughput even for compute-bound kernels by creating a power headroom for higher \gls{sm} frequencies through reduced data movement.
Extending this analysis beyond \glspl{gpu}, we find that power constraints also limit throughput scaling on Tenstorrent's Wormhole accelerator, highlighting the broader importance of accounting for physical non-idealities in kernel performance engineering.

\end{abstract}
]



\printAffiliationsAndNotice{}  

\glsresetall

\section{Introduction}


The fundamental idea of abstraction is pervasive across computer systems. By enabling the philosophy of \textit{separation of concerns}, abstractions have facilitated the development of software for increasingly complex hardware. 
At the highest level of abstraction, software is written in high-level languages and lowered to the target hardware with a compiler. This abstraction \textit{reduces the complexity of using hardware and improves productivity and portability \textup{\cite{chen2018tvm}}, but it also restricts the maximum performance achievable for the workload \textup{\cite{hagedorn2020fireiron}}.} As performance demands increase, performance engineers are compelled to break this level of abstraction and work with the hardware's \gls{isa} to develop kernels that effectively exploit the underlying microarchitecture \cite{hagedorn2020fireiron, shah2024flashattention}. Through this work, we show that even at this abstraction level, \textit{they are limited because physical non-idealities in \gls{ic} design, fabrication, and operation can cause the microarchitecture to behave differently under certain conditions.} 


\begin{figure}[tb]
    \centering
    \includegraphics[width=0.8\linewidth]{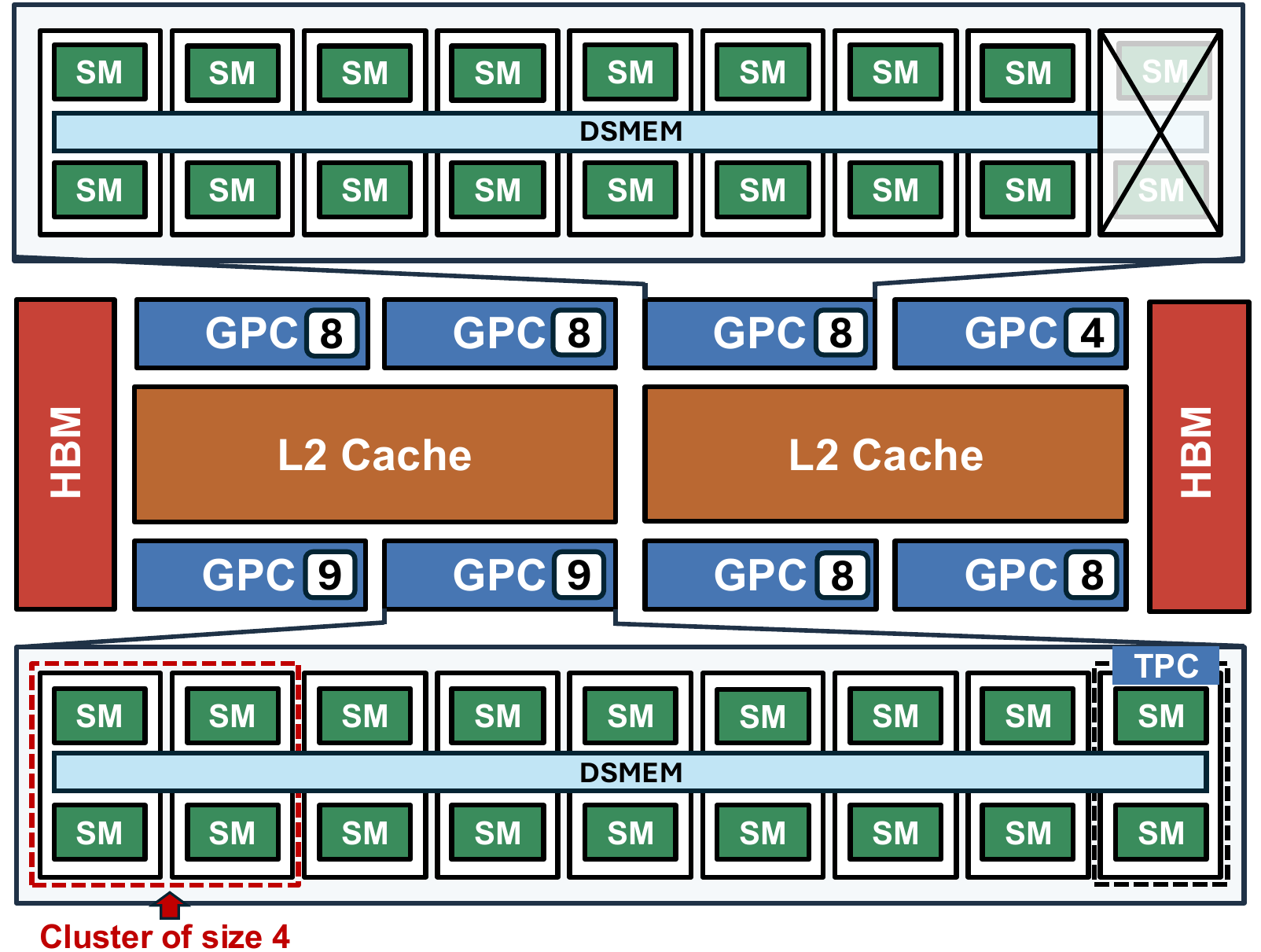}
    \caption{Layout of SMs in the NVIDIA H100 GPU. From our reverse engineering, the number of TPCs in each GPC is labelled.}
    \label{fig:gpu_architecture_figure}
\end{figure}

Physical non-idealities are physical constraints that cause \glspl{ic} to deviate from the expected behaviour that their microarchitecture may imply. Manufacturing defects that occur in \gls{ic} fabrication are one such constraint. Since it is common for modern \glspl{ic} to have defects with current manufacturing technologies, \gls{ic} vendors cannot restrict themselves to selling completely defect-free \glspl{ic}. As a solution, they disable the defective components and, if required, switch off additional functional components to align \glspl{ic} to a pre-determined resource profile. 
This technique is called \textit{floorsweeping}, and it allows the vendors to hide the distribution of faults and categorize a large percentage of faulty \glspl{ic} into a single product \cite{duluk2025virtualizing}. 
For example, \glspl{ic} manufactured with the NVIDIA GH100 design that have 144 \glspl{sm} are grouped into their enterprise data center-level \gls{gpu}, H100 SXM5, which has only 132 \glspl{sm} \cite{andersch2022hopper}. As users, we do not know how the 12 disabled \glspl{sm} are distributed in the \gls{gpu}. Therefore, a nominally identical \gls{gpu} model can have different internal physical topologies, even when its exposed compute capacity is identical. 
In addition to floorsweeping, \gls{tdp} specifies the maximum power an \gls{ic} is capable of consuming during a sustained workload. If a kernel attempts to use the microarchitecture in a manner that consumes more power than its \gls{tdp}, under nominal conditions, the \gls{ic} would throttle its frequency to reduce its power consumption \cite{nvidia2026benchmarking}. Since this is a dynamic process, its impact on kernel performance remains opaque at design time.
Physical non-idealities such as these are abstracted away by \gls{ic} vendors when they disclose their microarchitecture to provide consistent performance and programmability across a product line.

The impact these physical non-idealities have on kernel performance does not vary based on its design, in most cases. However, as kernel performance engineers try to realize the full performance potential of a microarchitecture, they may attempt optimizations that expose performance variations from these effects. In large-scale deployments like AI training and inference, such performance unpredictability can cause significant loss \cite{lin2025understanding}. We explore this issue in NVIDIA H100 \glspl{gpu} and demonstrate its impact with the \gls{dsmem} architectural feature, and show how the impact can be alleviated with kernel mappings that work around these non-idealities.

\begin{figure}[tb]
    \centering
    \includegraphics[width=\linewidth]{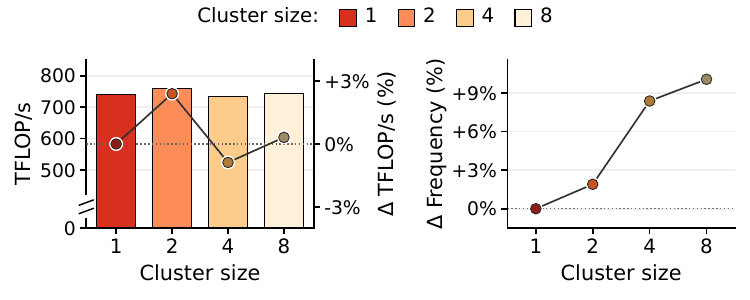}
    \caption{FlashAttention 4 throughput varies across cluster sizes due to physical non-idealities. The plot shows the throughput, percentage increase in throughput, and percentage increase in SM frequency compared to cluster size 1 for FlashAttention 4 with 64K sequence length and 256 head dimension.}
    \label{fig:fa4_seqlen_tflops_4k_16k_64k}
\end{figure}

NVIDIA GPUs before the Hopper architecture rely on sharing data between \glspl{sm} through the global memory, leading to congestion and consequent performance impacts \cite{andersch2022hopper}. The \gls{dsmem} introduced in the Hopper architecture reduces the inter-\gls{sm} communication cost via a \gls{noc} that connects shared memories of \glspl{sm} in a \gls{gpc}. As in Fig. \ref{fig:gpu_architecture_figure}, \glspl{sm} in the H100 \gls{gpu} are arranged into pairs called \glspl{tpc}, and nine of those \glspl{tpc} are grouped into a \gls{gpc} \cite{andersch2022hopper}. To utilize the \gls{dsmem}, multiple thread blocks need to be grouped, i.e., clustered, into a unit called a thread block cluster (cluster). It ensures that the thread blocks are mapped to \glspl{sm} within a single \gls{gpc}, as shown in Fig. \ref{fig:gpu_architecture_figure}. 
This architectural feature enables a thread block to multicast data from the global memory to all \glspl{sm}' shared memory in the cluster when they require the same data, a common scenario in tiled kernels \cite{nvidia2026multicast}. Interestingly, we observe that this feature is not completely utilized in modern kernels such as FlashAttention \cite{dao2022flashattention, shah2024flashattention, zadouri2026flashattention}. As seen in Fig. \ref{fig:fa4_seqlen_tflops_4k_16k_64k}, our evaluations on the Hopper version of FlashAttention 4 with multicast load also show that the expected performance gains do not materialize at cluster sizes 4 and 8. A kernel performance engineer might attribute this performance degradation entirely to synchronization overheads. 
However, our investigations uncover that it is only one of the contributing factors and that impacts from physical non-idealities dominate here.


In this paper, we first expose how floorsweeping limits the number of \glspl{sm} available for clustered execution. 
Although the NVIDIA H100 \gls{gpu} has 132 \glspl{sm} \cite{andersch2022hopper}, our experiments reveal that only 120 \glspl{sm} can be used concurrently at \gls{gpc}-level cluster sizes. 
We refer to this number as the active \gls{sm} limit. We explain this reduction using the organization of remaining functional resources into virtual \glspl{gpc}. A reduced active \gls{sm} limit decreases the number of concurrently scheduled thread blocks in the \gls{gpu}, i.e., the size of an execution wave, increasing the number of waves required for a kernel execution.
We characterize the conditions under which the reduced active \gls{sm} limit introduces additional execution waves. 
Further, we define a wave-equalized kernel launch as one whose thread-block count results in the same number of execution waves under both the full and reduced active \gls{sm} limits. 
With insights from this, we introduce the wave-aware clustering strategy that is cognizant of the floorsweeping effects. It is able to select the best clustering method for FlashAttention 4 with sequence lengths spanning from 2K to 64K. We achieve a throughput improvement of up to 5.30\% against the baseline FlashAttention 4 implementation, which uses cluster size 1. 

In contrast to floorsweeping, which causes a performance reduction on \gls{dsmem}, the \gls{tdp} limit has a positive impact. Surprisingly, we find that the throughput improvements from \gls{dsmem} do not always occur due to improved tensor core utilization from reduced data movement. In compute-bound kernels, as in Fig. \ref{fig:fa4_seqlen_tflops_4k_16k_64k}, the memory bandwidth optimization from \gls{dsmem} reduces the power consumed for data transfer, allowing the \gls{gpu} to sustain a higher \gls{sm} frequency with \gls{dvfs}. We achieve throughput improvements up to 3\% through this phenomenon. We extend this analysis beyond \glspl{gpu} to Tenstorrent's Wormhole accelerator and uncover that it also has a similar behaviour with the \gls{tdp} limit, where its expected throughput improvements with a higher number of active cores are restricted by power-limiting frequency variations. These results show that hardware non-idealities cause unexpected performance variations in kernels and that physical-constraint-aware kernel development can unlock meaningful performance improvements.

\section{Background and Related Work}
\vspace{-1mm}
\subsection{Distributed shared memory in GPUs}
\vspace{-1mm}
\Gls{dsmem} is an architectural feature introduced in the NVIDIA Hopper \gls{gpu} architecture. It connects the shared memories of multiple \glspl{sm} into a unified memory space through a \gls{noc} \cite{prabhakar2025distributed}. Thread blocks can be grouped into a unit called a thread block cluster (cluster), with supported cluster sizes of 2, 4, and 8, with 16 being optional. However, not all \glspl{sm} in the \gls{gpu} can be clustered together, as cluster placement is constrained by the physical organization of the \glspl{sm}. As shown in Fig. \ref{fig:gpu_architecture_figure}, the NVIDIA Hopper \gls{gpu} organizes \glspl{sm} into pairs within units called \glspl{tpc}. Nine \glspl{tpc} form a \gls{gpc}, and the \gls{gpu} contains eight such \glspl{gpc}. Each \gls{gpc} has a \gls{noc} connecting the shared memories of the \glspl{sm} within it. Consequently, only \glspl{sm} within the same \gls{gpc} can participate in a cluster. Furthermore, clusters of size 2 always group \glspl{sm} within a single \gls{tpc}. Therefore, we refer to clusters of size 2 as \gls{tpc}-level clusters and clusters of sizes 4 and 8 as \gls{gpc}-level clusters. While clustered execution has the potential to reduce data movement within the \gls{gpu}, it also introduces additional synchronization overhead and scheduling constraints. In this work, we use \gls{dsmem} to demonstrate that unexpected trade-offs arise from this  feature, which become apparent only when physical non-idealities are taken into account.

\subsection{Impact of physical non-idealities}
\vspace{-1mm}
Prior studies have reported that \gls{dsmem} throughput decreases with increasing \gls{gpc}-level cluster sizes \cite{xluo2026clusterfusion, wluo2025dissecting}. This behavior has been attributed to contention in the crossbar connecting the L1 caches/shared memories of \glspl{sm} within a \gls{gpc} \cite{xluo2026clusterfusion, prabhakar2025distributed}. Additionally, the number of \glspl{sm} available for clustered execution has been observed to decrease at \gls{gpc}-level cluster sizes \cite{xluo2026clusterfusion, hopper2026tuning}. Floorsweeping has been identified as a potential cause of these resource restrictions \cite{chen2026semianalysis, he2026taming}. Concurrent work by Fan et al. \cite{fan2026dissecting} characterizes the physical \gls{sm} topology and examines its implications for thread-block cluster scheduling, particularly in the context of MIG partitioning \cite{nvidia_mig} and Green Contexts \cite{nvidia_green_contexts}.
Beyond the resource restrictions imposed by floorsweeping, the performance of clustered kernels also depends on the operating frequency sustained within the available power budget. Existing Hopper microbenchmarking studies have characterized \gls{dsmem} latency and throughput and separately reported power-induced frequency reductions during tensor-core execution. Studies of \gls{gpu} power capping and frequency scaling have also characterized performance and energy trade-offs, but have not isolated how \gls{dsmem}-based reductions in data movement affect operating frequency \cite{ujeniya2026architectural, afzal2026modeling}. More broadly, the effects of manufacturing variations and power constraints on CPU and \gls{gpu} performance have been studied at the microarchitectural level \cite{lee2011improving, rountree2012beyond, aguilera2014process, schoonhoven2022going}.
\textit{In this work, we first characterize the restrictions imposed by floorsweeping on clustered execution and its performance implications. We propose the conditions for kernel launches that avoid additional execution waves under these restrictions. Furthermore, we investigate how \gls{tdp}-induced frequency scaling affects clustered kernel performance and explain its combined impact with floorsweeping.}
\section{Methodology}
\vspace{-2mm}
We evaluate four NVIDIA H100 SXM5 \glspl{gpu} to assess the impact of physical non-idealities and ensure that the observed effects are not specific to a single \gls{gpu}. The four \glspl{gpu} are located in geographically distinct regions to reduce the likelihood that the observations are specific to a particular manufacturing batch. Throughout this paper, NVIDIA H100 refers to the NVIDIA H100 SXM5 \gls{sku} unless otherwise specified. We use four CUDA kernels for our experiments: a multicast load microbenchmark, a \gls{gpu} skyline reverse-engineering kernel, GEMM, and FlashAttention. All kernels are compiled using CUDA 13.0 with the \texttt{sm\_90a} compilation target. Each kernel is executed for multiple warm-up and timed iterations, with the exact counts reported in the corresponding experimental sections. Furthermore, we record the SM clock frequency using the NVIDIA Management Library (NVML) and the remaining profiling data are collected using \gls{ncu}.

\begin{figure*}[tb]
    \centering
    \subfigure[Single-block multicast load pattern.]{%
        \includegraphics[width=0.45\linewidth]{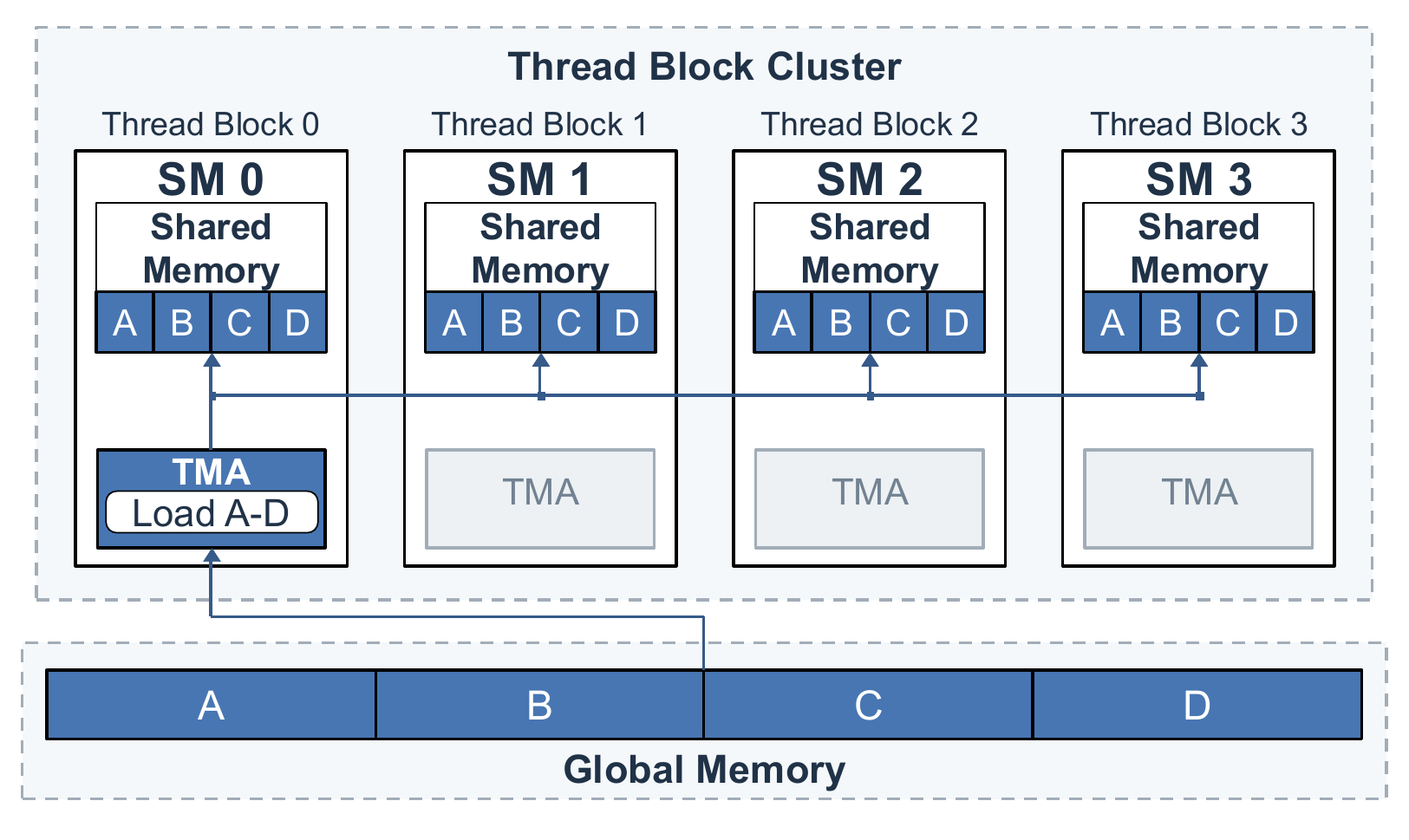}%
        \label{fig:single_block_broadcast_pattern}%
    }\hfill%
    \subfigure[All-blocks multicast load pattern.]{%
        \includegraphics[width=0.45\linewidth]{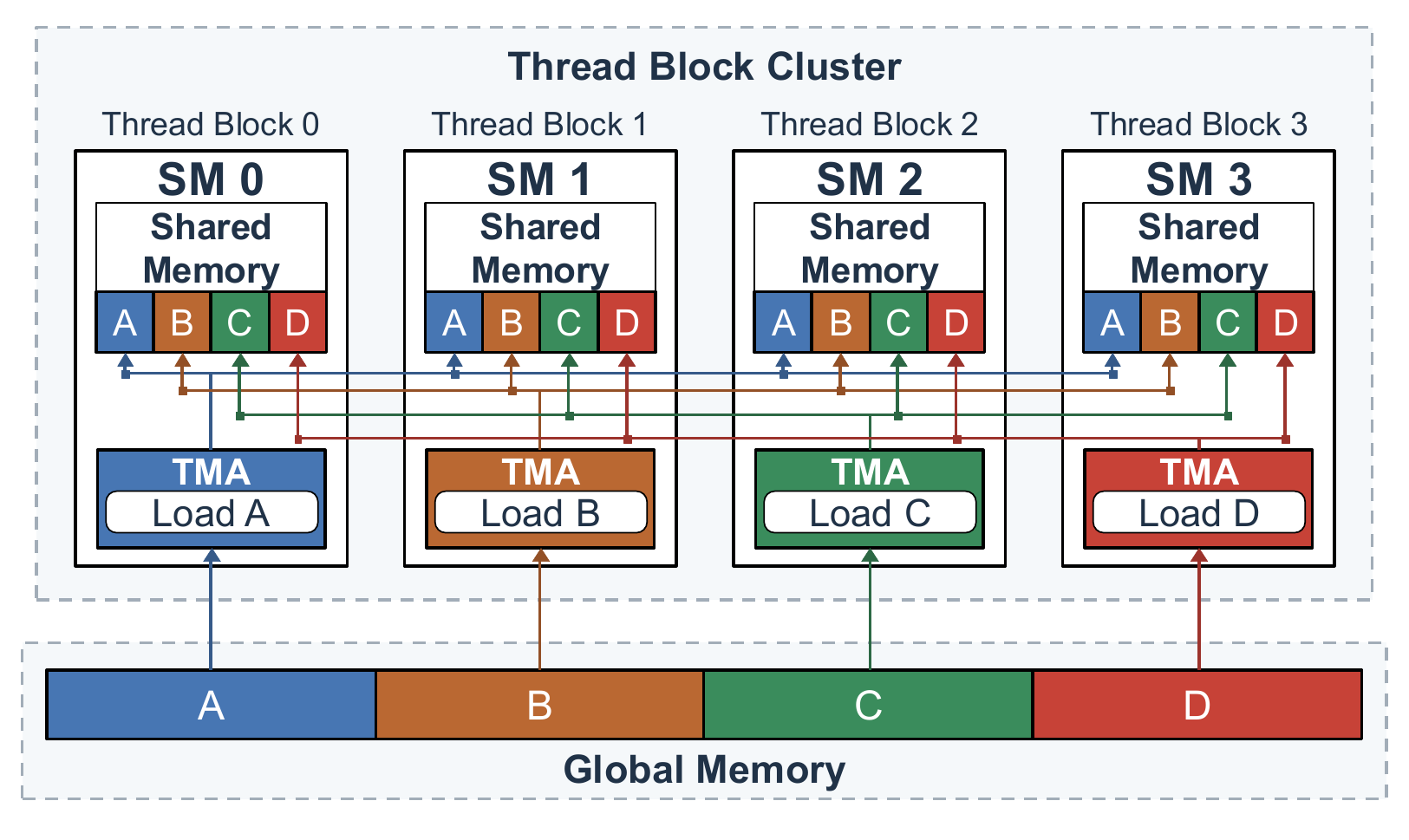}%
        \label{fig:all_blocks_broadcast_pattern}%
    }
    \caption{Multicast load patterns visualized for cluster size 4.}
    \label{fig:gload_broadcast_patterns}
\end{figure*}

For GEMM, we use the Hopper warp-specialized GEMM kernel from the CUTLASS 4.8.0 library \cite{cutlass_hopper_gemm}. We configure it with three pipeline stages and use a non-persistent execution mode to ensure that, for a fixed problem size, the number of thread blocks depends only on the tile dimensions. For FlashAttention, we use the FlashAttention 4 Hopper forward kernel \cite{flashattention4sm90}. The kernel takes query ($Q$), key ($K$), and value ($V$) matrices as inputs. With a tiling configuration of $B_M \times B_N$, each thread block processes $B_M$ rows of $Q$ and iteratively loads tiles of $B_N$ rows from $K$ and $V$ to perform the attention operation. Since multiple thread blocks processing different query tiles require the same $K$ and $V$ tiles, there is an opportunity to reduce redundant data loads through multicast loads using \gls{dsmem}. We modify the FlashAttention codebase to incorporate this functionality and evaluate it using non-causal multi-head attention (MHA).

For the Tenstorrent experiments, we use one Wormhole chip of an n300d card and TTNN~\cite{tenstorrent_ttnn}, Tenstorrent's official library built on the TT-Metalium software stack~\cite{tenstorrent_ttmetal}. We run matrix multiplication with \texttt{ttnn.matmul} on BF16 inputs, using the official TTNN GEMM benchmark with its default configuration. We vary the number of active Tensix cores from 1 to 64 and keep the work per core fixed. While the benchmark runs, a separate process reads the chip's core clock  and power from the firmware telemetry that the Tenstorrent kernel driver exposes in sysfs. We report the median of the samples taken during each measurement.
\section{Linking Cluster Scaling Limits to Floorsweeping}
\label{sec:gload_and_bcast_in_cluster}
\vspace{-1mm}
Distributed shared memory enables a thread block to initiate a multicast load from global memory into the shared memories of the corresponding \glspl{sm} within a cluster, eliminating redundant data loads. This is supported by the network connecting the shared memories of \glspl{sm} within a \gls{gpc} \cite{prabhakar2025distributed}. When thread blocks are grouped into a cluster, they are scheduled on \glspl{sm} within the same \gls{gpc}. Such data sharing can reduce L2 cache traffic and the cumulative data loading time across thread blocks. However, these expected benefits do not consistently translate into improved kernel performance, as shown in Fig. \ref{fig:fa4_seqlen_tflops_4k_16k_64k}. To investigate the underlying causes, we develop a microbenchmark to isolate data loading performance.

\subsection{Microbenchmarking multicast loads}
\vspace{-1mm}
In this microbenchmark, the CPU generates 1 MB of pseudorandom FP32 values, which are then copied to the \gls{gpu}'s global memory (HBM). We launch the kernel with 256 thread blocks and evaluate cluster sizes of 1, 2, 4, and 8 in separate runs. For cluster sizes greater than 1, the kernel performs multicast loads of the data in successive 128 KB chunks using the two patterns illustrated in Fig. \ref{fig:gload_broadcast_patterns}. In the single-block multicast load pattern, the first thread block in each cluster initiates a \gls{tma} transfer to load the data into the shared memory of every thread block within the cluster. In contrast, in the all-blocks multicast load pattern, each thread block initiates a \gls{tma} transfer for a distinct $1/\text{(cluster size)}$ portion of the data, multicasting its portion into the shared memory of the corresponding \glspl{sm} of each thread block within the cluster.

\begin{figure}[htb]
    \centering
    \includegraphics[width=0.8\linewidth]{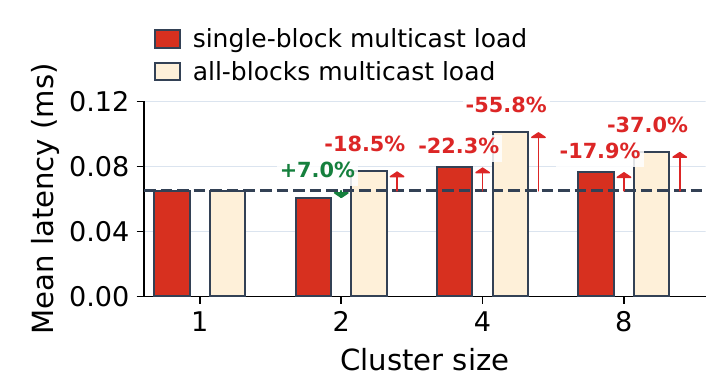}
    \caption{Mean latency for 256 thread blocks to load 1 MB of data in 128 KB chunks from global memory into shared memory using multicasting, across different cluster sizes.}
    \label{fig:gload_broadcast_1024KiB_latency}
\end{figure}

The kernel is executed for 50 warm-up iterations, followed by 500 timed iterations to calculate the mean latency. Furthermore, the \gls{gpu}'s \gls{sm} and memory clock frequencies are fixed at their maximum values of 1980 MHz and 2619 MHz, respectively. At cluster sizes greater than 1, the all-blocks multicast load pattern issues the same number of \gls{tma} transfers as the cluster size 1 configuration, but with smaller transfer sizes than the single-block multicast load pattern. Since smaller \gls{tma} transfers can reduce effective memory bandwidth \cite{wluo2025dissecting}, the single-block multicast load pattern achieves lower latency than the all-blocks multicast load pattern, as shown in Fig. \ref{fig:gload_broadcast_1024KiB_latency}.

The microbenchmark results also reveal an unexpected trend at \gls{gpc}-level cluster sizes, similar to that observed in Fig. \ref{fig:fa4_seqlen_tflops_4k_16k_64k}. Since \gls{gpc}-level cluster sizes progressively reduce the amount of data transferred from the L2 cache to shared memory, they are expected to achieve lower mean latency than the \gls{tpc}-level cluster size. However, this expected improvement does not materialize due to the reduced number of active \glspl{sm} available for \gls{gpc}-level clustered execution.

\subsection{Active SM limits at different cluster sizes}
\label{sec:active_sm_limits_at_diff_cluster_sizes}
\vspace{-1mm}
To determine the number of active \glspl{sm} at different cluster sizes, we record the logical \gls{sm} IDs used by the microbenchmark kernel during each iteration. The logical \gls{sm} ID is obtained by having the first thread of each thread block read the \textit{\%smid} register, which uniquely identifies the \gls{sm} executing the thread block. Table \ref{tab:unique_smids_at_cluster_sizes} reports the number of unique \glspl{sm} observed within individual iterations and across all iterations for each cluster size.

\begin{table}[htb]
\caption{Number of unique active \glspl{sm} at each cluster size.}
\label{tab:unique_smids_at_cluster_sizes}
\centering
\small
\renewcommand{\arraystretch}{0.75}
\begin{tabular}{lcc}
\toprule
& \multicolumn{2}{c}{Unique \glspl{sm}} \\
\cmidrule(lr){2-3}
Cluster Size & Per Iter. & All Iters. \\
\midrule
1 & 132 & 132 \\
2 & 132 & 132 \\
4 & 120 & 124 \\
8 & 120 & 124 \\
\bottomrule
\end{tabular}
\end{table}

At cluster sizes 1 and 2, the kernel utilizes all 132 \glspl{sm} in the NVIDIA H100 \gls{gpu}. However, at \gls{gpc}-level cluster sizes, only 124 \glspl{sm} are enabled for clustered execution, of which at most 120 can be used concurrently. Surprisingly, these restrictions arise from floorsweeping.
\subsection{Effect of Floorsweeping on active SM limit}
\vspace{-1mm}
Physical non-idealities can cause \glspl{ic} to differ from the resource organization implied by their microarchitectural design. For large dies, manufacturing defects are one source of deviation affecting a substantial fraction of chips. The negative binomial (NB) model expresses the zero-defect yield as $Y=\left(\frac{\alpha}{\alpha+\lambda_0}\right)^\alpha$, where $\alpha$ controls the variation in defect counts across chips. The expected number of defects per chip is $\lambda_0=D_0 A$, where $D_0$ is the defect density and $A$ is the die area~\cite{yieldmodel, statistical-yield-modeling, supplychain}. At a defect density of $0.15$ defects/cm\textsuperscript{2}, a large accelerator die near the reticle limit ($700$--$800$ mm\textsuperscript{2}) has approximately $1.05$ defects on average. Smaller $\alpha$ values represent greater clustering, where some chips contain substantially more defects than others. Requiring every chip to be defect-free or have identical defect patterns is impractical. Vendors therefore disable defective components through \textit{floorsweeping} to expose a usable resource profile. Combined with \textit{binning}, this allows physically different chips to be sold within the same SKU, while chips with fewer surviving resources may qualify for lower-performance SKUs.

For example, the NVIDIA GH100 design contains 144 \glspl{sm}, whereas the H100 SXM5 exposes 132. This resource count does not reveal which 12 \glspl{sm} are disabled or their distribution across the chip. Consequently, GPUs of the same SKU can expose identical compute capacity but different internal physical topologies. Assuming a defect disables a TPC, some physical GPCs may lose more TPCs than others, resulting in different distributions of active \glspl{sm} per GPC. NVIDIA's patent~\cite{duluk2025virtualizing} introduces a \textit{Skyline}, specifying a common distribution of TPCs across virtual GPCs (vGPCs) for a SKU. Each tower represents a vGPC, with its height specifying the number of exposed TPCs.

\begin{figure}[tb]
    \centering
    \includegraphics[width=0.75\linewidth]{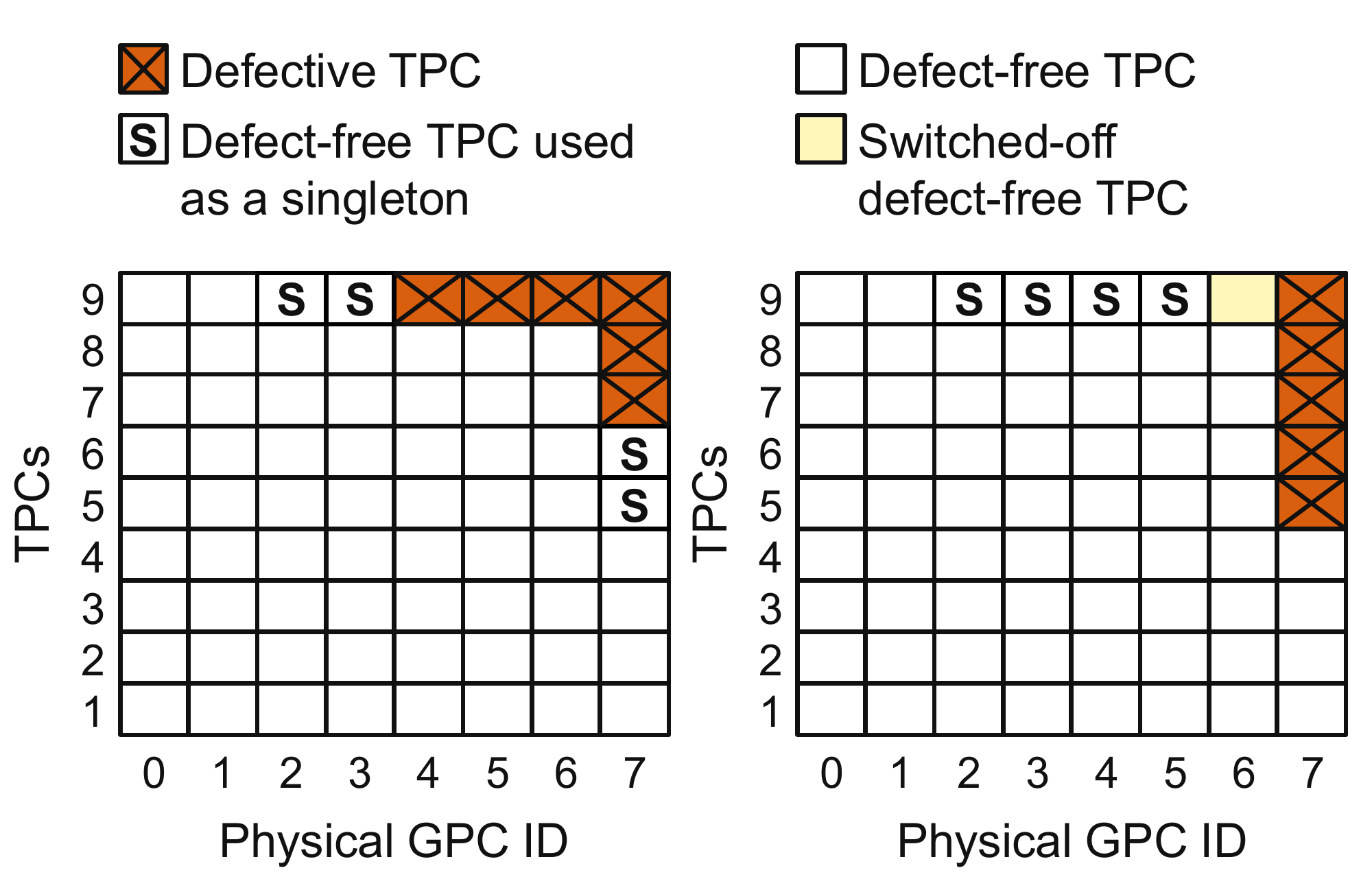}
    \caption{Hypothetical fault patterns mapped to a common skyline through floorsweeping~\cite{duluk2025virtualizing}.}
    \label{fig:skylines}
\end{figure}

Figure~\ref{fig:skylines} illustrates how dies with different physical fault patterns can expose the same skyline. Each vGPC is formed from surviving TPCs within a physical GPC. A physical GPC with more surviving TPCs than required by its mapped tower can expose the remaining TPCs separately as single-TPC vGPCs, called \textit{singletons}. Thus, tower heights describe the virtual grouping available to software rather than the physical distribution of surviving resources. Skyline selection aims to accommodate as many physical configurations as possible while minimizing resource loss.

For example, a skyline of $6/8/8/8/9/9/9/9/1{\times}2$ exposes 68 TPCs across eight multi-TPC vGPCs and two singletons.  \textit{Singletons} help accommodate different fault patterns but also constrain clustered execution. They can execute independent thread blocks or TPC-level clusters, but cannot participate in multi-TPC (GPC-level) clusters. Thus, accommodating more configurations may require disabling otherwise functional TPCs or exposing them in smaller virtual groups, limiting the applicability and effectiveness of clustered execution.

\subsection{Reverse engineering GPU skylines and its impact}
\vspace{-1mm}
The earlier microbenchmark records only 120 unique \gls{sm} IDs per iteration at cluster sizes 4 and 8, and 124 across all iterations, despite the H100 exposing 132 \glspl{sm} (Table~\ref{tab:unique_smids_at_cluster_sizes}). Since the \glspl{sm} in a cluster must belong to the same vGPC, both the sizes of these groups and the clusters they can accommodate can fragment available compute capacity. To understand this discrepancy, we reverse-engineer the H100 skyline from the observed \gls{sm} groupings during clustered execution.

We launch a CUDA kernel with 1024 thread blocks of 1024 threads each. The first thread of each block reads the \texttt{\%smid} register and stores its ID in global memory. For increasing cluster sizes, we group the \gls{sm} IDs belonging to each cluster into sets, leveraging the grouping of consecutive thread blocks during clustering. We merge sets sharing one or more \gls{sm} IDs to infer the underlying \gls{sm} groupings, repeating the process for 1000 iterations to account for variations.

We infer 12 vGPCs with the skyline $9/9/8/8/8/8/8/4/1/1/1/1$, as shown in Table~\ref{tab:skyline_h100}. We assign vGPC IDs in ascending order of the lowest \gls{sm} ID within each group. This virtual organization contains 12 groups despite the chip having only eight physical GPCs.
The four-TPC vGPC potentially exists to accommodate dies with faults largely concentrated within one physical GPC.

\begin{table}[t]
\caption{Inferred skyline for the measured NVIDIA H100.}
\label{tab:skyline_h100}
\vskip 0.15in
\begin{center}
\begin{small}
\renewcommand{\arraystretch}{0.75}
\begin{tabular}{lrr}
\toprule
Type & Count & vGPC IDs\\
\midrule
2 \glspl{sm} vGPC (1 TPC) & 4 & 8,9,10,11\\
8 \glspl{sm} vGPC (4 TPCs) & 1 & 0\\
16 \glspl{sm} vGPC (8 TPCs) & 5 & 1,2,3,4,5\\
18 \glspl{sm} vGPC (9 TPCs) & 2 & 6,7\\
\bottomrule
\end{tabular}
\end{small}
\end{center}
\end{table}

This skyline explains both observed \gls{sm} counts. The four singleton vGPCs contain two \glspl{sm} each and cannot support clusters of size 4 or 8, leaving only 124 of the 132 \glspl{sm} available for GPC-level clustered execution. Furthermore, each of the two 18-\gls{sm} vGPCs leaves two \glspl{sm} unassigned when partitioned into clusters of size 4 or 8. This further reduces the active \gls{sm} limit to 120 for these cluster sizes in the measured configuration. Since the unassigned \glspl{sm} can vary between iterations, the microbenchmark observes 120 unique \gls{sm} IDs per iteration but 124 across all iterations.

\begin{figure}[tb]
    \centering
    \includegraphics[width=0.8\linewidth]{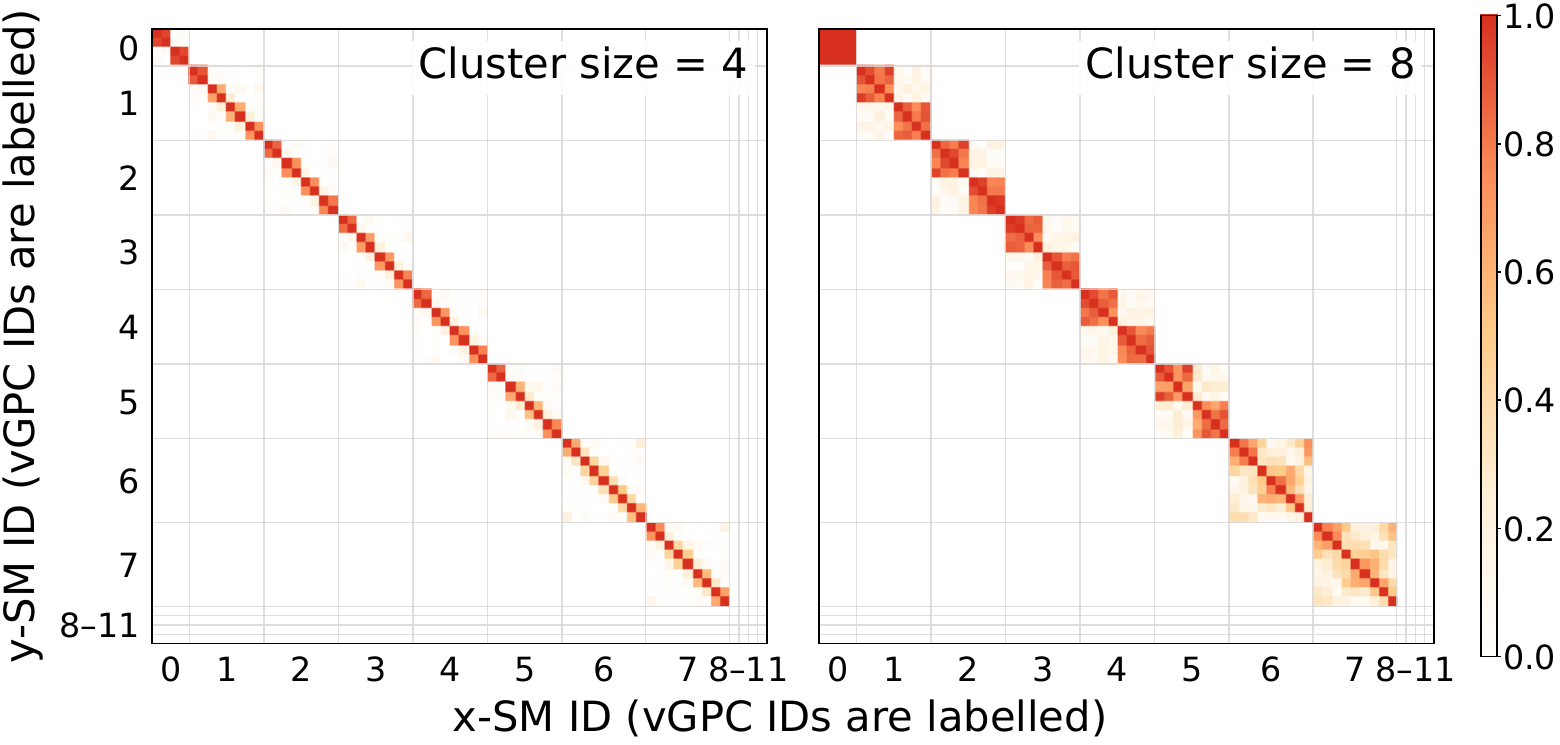}
    \caption{Probability that SM $y$ belongs to a cluster given that SM $x$ belongs to the same cluster, for cluster sizes 4 and 8.}
    \label{fig:smid_heatmap_cluster_4}
\end{figure}

Beyond restricting the availability of \glspl{sm}, the underlying resource organization also influences which \glspl{sm} are frequently grouped into clusters. As shown in Figure~\ref{fig:smid_heatmap_cluster_4}, \glspl{sm} with neighboring logical \gls{sm} IDs are frequently clustered together at cluster sizes 4 and 8. Thus, a performance engineer must account for both the number of available \glspl{sm} and how they are grouped for clustered execution. The reduction in the active \gls{sm} limit from 132 to 120 can increase the number of execution waves and reduce kernel performance. The following section characterizes when this additional-wave penalty occurs and identifies configurations in which the execution wave count remains unchanged.

\section{Navigating Active SM Limit Penalties}
\label{sec:nav_sm_limit_penalty}
\vspace{-1mm}
The active \gls{sm} limit imposed by floorsweeping can increase the number of execution waves required by a kernel, as the same number of thread blocks must be scheduled across fewer \glspl{sm}. An execution wave consists of the thread blocks that can execute concurrently on the \gls{gpu}. The number of execution waves influences kernel execution time, as additional waves require successive scheduling of thread blocks. Consequently, the additional-wave penalty caused by the reduced active \gls{sm} limit can be avoided when a kernel launch requires the same number of execution waves under both the full 132-\gls{sm} and reduced 120-\gls{sm} limits. We refer to such a launch as a wave-equalized kernel launch, formalized in Eq. \ref{eq:wave_equalized_kernel_launch_condition}. 
We see pervasive LLM kernels such as GEMM and FlashAttention often achieve their highest throughput by maximizing data reuse through large tile shapes that fit within the available shared memory and register capacity. The resulting resource requirements typically limit occupancy to one thread block per SM, and this restricted occupancy exacerbates the impact of the reduced active \gls{sm} limit because it minimizes the number of thread blocks in a wave.

\begin{equation}
\label{eq:wave_equalized_kernel_launch_condition}
\lceil\frac{\text{\# Thread Blocks}}{132}\rceil = \lceil\frac{\text{\# Thread Blocks}}{120}\rceil
\end{equation}

\subsection{Wave-equalized multicast load within clusters}
\begin{figure}[tb]
    \centering
    \begin{minipage}[t]{0.48\linewidth}
        \vspace{0pt}
        \centering
        \includegraphics[width=\linewidth]{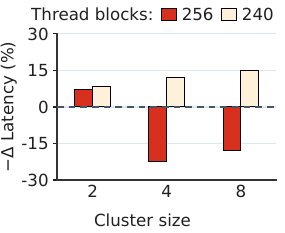}
        \caption{Reduction in latency of loading 1 MB data (128 KB chunks) from global memory to shared memories with multicast across cluster sizes.}
        \label{fig:wave_equalized_single_block_broadcast_1024KiB_latency}
    \end{minipage}\hfill
    \begin{minipage}[t]{0.48\linewidth}
        \vspace{0pt}
        \centering
        \includegraphics[width=\linewidth]{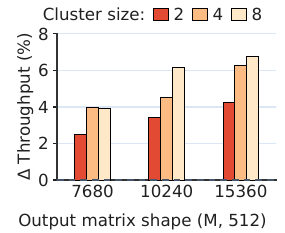}
        \caption{\(M \times 512 \times 4096\) GEMM throughput improvement relative to cluster 1. $M$ (x-axis) is varied across wave-equalized kernel launches.}
        \label{fig:gemm_m_sweep_n512_k4096_throughput_improvement}
    \end{minipage}
\end{figure}

\vspace{-1mm}
To examine how wave equalization influences the benefits of multicast load, we compare the original 256-thread-block microbenchmark from Section \ref{sec:gload_and_bcast_in_cluster} with a 240-thread-block wave-equalized configuration satisfying Eq. \ref{eq:wave_equalized_kernel_launch_condition}. Figure \ref{fig:wave_equalized_single_block_broadcast_1024KiB_latency} shows the change in latency due to clustering, relative to the cluster size 1 baseline, for both configurations. At \gls{gpc}-level cluster sizes, clustering reduces latency in the wave-equalized 240-block configuration but increases it in the non-wave-equalized 256-block configuration. These results illustrate how additional execution waves can offset the latency benefits of reduced data movement through clustering.

\subsection{Wave-equalized GEMM kernel launches}
\vspace{-1mm}
To evaluate the effect of wave equalization, we first characterize GEMM performance across different cluster sizes. 
We evaluate tile shapes of \(128 \times 128\) and \(64 \times 64\), along with all valid cluster shapes for cluster sizes greater than 1. For each GEMM problem shape and cluster size, we select the configuration that achieves the highest throughput. Furthermore, each configuration is executed for 100 warm-up iterations, followed by 1000 timed iterations.  

As shown in Table \ref{tab:gemm_throughput_and_hbmutil}, the L2 hit rate of the unclustered kernel decreases as the operand size increases, because a smaller proportion of the operand data gets reused through the L2 cache. This causes a rise in HBM reads, causing the HBM utilization to increase to near saturation. When the HBM is highly utilized, the bandwidth optimization provided by clustering outweighs its synchronization overheads and reduced active \gls{sm} limit penalties. Conversely, the clustering bandwidth optimization has no effect when the HBM utilization is low. Interestingly, the impact of reduced active \gls{sm} limit is pronounced at the middle of these extremes, where the memory bandwidth optimization seems to overpower the synchronization overheads but fails to compensate for the additional execution waves at \gls{gpc}-level clustering. To examine the impact of working around this additional wave penalty, we evaluate wave-equalized GEMM configurations.




\begin{table}[htb]
\caption{HBM utilization and the  improvement in throughput over cluster 1, for GEMM (tile size $64 \times 64$).}
\label{tab:gemm_throughput_and_hbmutil}
\centering
\footnotesize
\resizebox{\linewidth}{!}{
\begin{tabular}{llcccc}
\toprule
& & \multicolumn{4}{c}{Cluster size} \\
\cmidrule(lr){3-6}
M $\times$ N $\times$ K & Metric & 1 & 2 & 4 & 8 \\
\midrule
\multirow{3}{*}{4K $\times$ 4K $\times$ 4K}
& $\Delta$ Throughput (\%) & - & -9.6\% & -18.8\% & -21.4\% \\
& HBM Util. (\%) & 22.9\% & 21.8\% & 21.0\% & 14.7\% \\
& L2 Hit Rate (\%) & 89.2\% & 85.8\% & 82.7\% & 86.5\% \\
\midrule
\multirow{3}{*}{8K $\times$ 8K $\times$ 4K}
& $\Delta$ Throughput (\%) & - & +6.1\% & +0.4\% & -2.2\% \\
& HBM Util. (\%) & 61.5\% & 36.3\% & 32.0\% & 17.9\% \\
& L2 Hit Rate (\%) & 74.1\% & 79.9\% & 79.6\% & 89.0\% \\
\midrule
\multirow{3}{*}{32K $\times$ 32K $\times$ 4K}
& $\Delta$ Throughput (\%) & - & +32.0\% & +41.2\% & +43.2\% \\
& HBM Util. (\%) & 85.8\% & 58.5\% & 33.1\% & 18.9\% \\
& L2 Hit Rate (\%) & 60.2\% & 71.7\% & 82.0\% & 89.5\% \\
\bottomrule
\end{tabular}
}
\end{table}



For a GEMM kernel launch to be wave-equalized, its output dimensions and tile shape must satisfy \(\lceil\frac{\lceil\frac{M}{B_M}\rceil \cdot \lceil\frac{N}{B_N}\rceil}{132}\rceil = \lceil\frac{\lceil\frac{M}{B_M}\rceil \cdot \lceil\frac{N}{B_N}\rceil}{120}\rceil\), where $M \times N$ denotes the output matrix dimensions and $B_M \times B_N$ denotes the tile shape. We fix $N$ and $K$ of GEMM to 512 and 4096, respectively, and vary $M$ to satisfy this condition for the $128 \times 128$ tile shape while being eligible for all cluster sizes. Figure \ref{fig:gemm_m_sweep_n512_k4096_throughput_improvement} shows that \gls{gpc}-level clustering improves throughput relative to cluster size 1 for these wave-equalized workloads. 




\subsection{Wave-equalized FlashAttention kernel launches}
\label{sec:wave-equalized flash attention kernel launches}
\vspace{-1mm}
For a FlashAttention kernel launch to be wave-equalized, it must satisfy Eq. \ref{eq:wave_equalized_kernel_launch_condition}. The number of thread blocks in a FlashAttention kernel is given by $B \times H \times \lceil \frac{S}{B_M} \rceil$, where $B$ is the batch size, $H$ is the number of attention heads, $S$ is the sequence length, and $B_M$ is the tile size along the query sequence dimension. Accordingly, the wave-equalization condition for FlashAttention can be expressed as \(\lceil \frac{B \times H \times \lceil\frac{S}{B_M}\rceil}{132}\rceil = \lceil \frac{B \times H \times \lceil\frac{S}{B_M}\rceil}{120}\rceil\). Since each batch and attention head has its own $K$ and $V$ matrices, clustering is performed independently for each batch and head. Therefore, to support cluster sizes of up to 8, the additional condition $\lceil \frac{S}{B_M} \rceil \bmod 8 = 0$ must also be satisfied.


\begin{figure}[tb]
    \centering
    \includegraphics[width=0.8\linewidth]{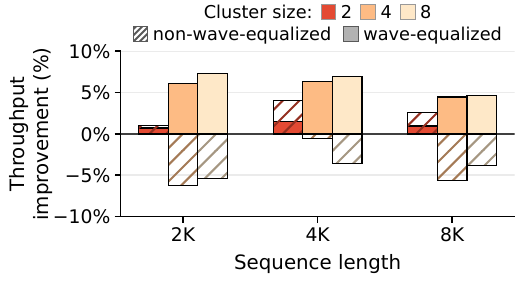}
    \caption{Percentage increase in throughput over cluster size 1 across sequence lengths for wave-equalized and non-wave-equalized FlashAttention 4 kernel launches.}
    \label{fig:cluster_improvement_hdim256_bars_16_heads_vs_15_heads}
\end{figure}

\begin{figure*}[tb]
    \centering
    \includegraphics[width=0.8\linewidth]{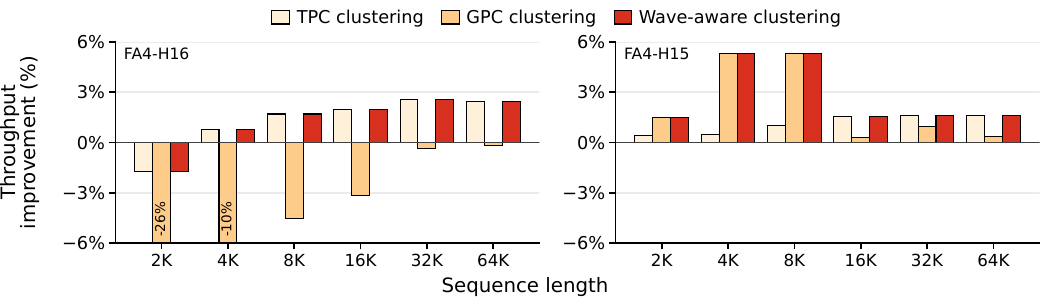}
    \caption{Percentage throughput improvement over FlashAttention-4 (uses cluster size 1) for fixed TPC/GPC clustering and non-ideality-informed wave-aware clustering. Wave-awareness matches the better of the two without requiring ad hoc profiling for each shape.}
    \label{fig:cluster_strategy_improvement_hdim256}
\end{figure*}

To examine how clustering performs across workloads with different execution wave counts, we fix the total number of tokens at 8192, setting the batch size to $8192/S$ for sequence length $S$. We fix the head dimension at 256 and evaluate workloads with 16 and 15 attention heads, representing non-wave-equalized and wave-equalized configurations, respectively. Each clustered configuration is compared against the cluster size 1 baseline for the same workload. We evaluate all combinations of $B_M \in \{128, 64, 32\}$ and $B_N \in \{80, 64, 32\}$, selecting the tile shape with the highest throughput. The maximum $B_N$ is limited to 80 because a $128 \times 128$ tile exceeds the available shared memory capacity at a head dimension of 256. Each configuration is executed for 100 warm-up iterations, followed by 1000 timed iterations. Across all configurations, the $128 \times 80$ tile achieves the highest throughput. With $B_M = 128$, the 15-head configurations satisfy the wave-equalization condition for FlashAttention. Figure \ref{fig:cluster_improvement_hdim256_bars_16_heads_vs_15_heads} shows that \gls{gpc}-level clustering improves throughput over cluster size 1 across wave-equalized configurations, with gains up to 7.32\%, exceeding those in non-wave-equalized configurations by up to 12.69 percentage points. Based on this characterization, we propose \textit{wave-aware clustering}, a strategy that chooses between TPC and GPC clustering for a given kernel based on its number of execution waves.

\subsection{Wave-aware clustering on FlashAttention}
\vspace{-1mm}
FlashAttention 4 kernel made for NVIDIA Hopper \glspl{gpu} does not use clustering, i.e., uses cluster size 1. We suspect that it is due to the performance variability in clustered execution that is unexplainable from a microarchitectural perspective. As the impact of floorsweeping sheds light on this variation in kernel performance, we propose the \textit{wave-aware clustering strategy} in Eq. \ref{eq:wave_aware_clustering_strategy}, where TB denotes thread blocks, for choosing between \gls{tpc}-level and \gls{gpc}-level clustering. To evaluate it, we consider FlashAttention 4 for varying sequence lengths from 2K to 64K, as we see in Fig. \ref{fig:cluster_strategy_improvement_hdim256}, under two configurations: FA4-H16 and FA4-H15. Both configurations have 256 head dimension and a batch size of 1, while their number of heads are 16 and 15, respectively. As in the previous FlashAttention experiment, the $128 \times 80$ tile size gave the highest throughput for each configuration, and they are executed for 100 warm-up iterations, followed by 100 timed iterations to measure kernel performance. Moreover, the TPC and GPC clustering conditions represent typical approaches driven by heuristics and profiling of expected workload mix, unaware of the true nature of the non-idealities.

\begin{equation}
\label{eq:wave_aware_clustering_strategy}
\lceil\frac{\text{\# TB}}{120}\rceil - \lceil\frac{\text{\# TB}}{132}\rceil = 
\begin{cases} 
      0 & => \text{GPC clustering} \\
      >0 & => \text{TPC clustering} 
\end{cases}
\end{equation}

Wave-aware clustering performs better than the FlashAttention 4 baseline that uses cluster size 1 at almost all evaluated kernel configurations and has throughput improvements up to 5.30\%. Further, it outperforms strictly or ad hoc using cluster size 2 (TPC clustering) or choosing cluster sizes 4 and 8 (GPC clustering) at all evaluated instances, as shown in Fig. \ref{fig:cluster_strategy_improvement_hdim256}. The throughput improvements observed with clustering in FlashAttention here are not driven by increased tensor core utilization resulting from reduced data movement. Counterintuitively, these improvements arise from the \gls{tdp} limit imposed during \gls{ic} design and its impact on operating frequency. We explain this further in Section \ref{sec:impact_of_thermal_design_power}.

Characterization of wave-equalized kernel launches and wave-aware clustering will help kernel performance engineers interpret performance variations and assess whether clustering is beneficial for a given workload and launch configuration. Moreover, it can also guide the selection of alternative launch geometries where workload constraints permit. These findings further reinforce the importance of looking beyond conventional microarchitectural abstractions and accounting for physical non-idealities in \gls{ic} fabrication when engineering high-performance kernels.
\section{Demystifying the Impact of Thermal Design Power}
\label{sec:impact_of_thermal_design_power}
\vspace{-1mm}
\begin{figure}[tb]
    \centering
    \includegraphics[width=\linewidth]{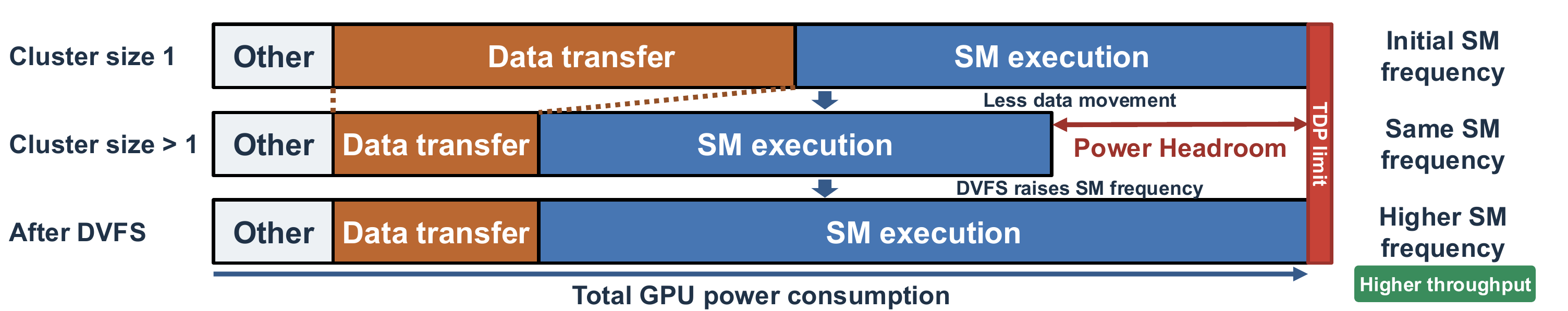}
    \caption{Effect of reduced data movement in a kernel on GPU power allocation under a fixed TDP limit.}
    \label{fig:tdp_dvfs_headroom}
\end{figure}

Thermal design power (TDP) specifies the heat a cooling system must dissipate during sustained operation. The NVIDIA H100 \gls{gpu} has a \gls{tdp} of 700 W, which is its approximate sustained power budget. Power consumption consists of static power, primarily from transistor leakage currents, and dynamic power, resulting from transistor switching activity. Dynamic power typically dominates compute-intensive workloads and depends on switching activity, operating frequency, and supply voltage. To regulate power consumption, \gls{dvfs} dynamically adjusts the operating voltage and frequency based on workload demands and power constraints. During low utilization, \gls{dvfs} reduces voltage and frequency to save power, while higher utilization allows the \gls{gpu} to increase its operating frequency, including that of the \glspl{sm}, provided sufficient power headroom is available. However, as power consumption approaches the \gls{tdp}-associated power limit, \gls{dvfs} may constrain or reduce operating frequency to remain within the permitted budget.

\subsection{Effect of TDP on kernel performance}
\vspace{-1mm}
To evaluate the impact of \gls{tdp} on kernel performance, we execute FlashAttention kernels with a fixed total token count of 256K, varying the batch size with sequence length. Furthermore, we maintain a constant hidden dimension of 2048 by adjusting the number of attention heads as the head dimension varies. As in the previous FlashAttention experiment, we evaluate all combinations of \(B_M \in \{128, 64, 32\}\) and \(B_N \in \{80, 64, 32\}\), selecting the tile shape that achieves the highest throughput for each configuration. Each configuration is executed for 100 warm-up iterations, followed by 100 timed iterations to measure kernel performance.

\begin{figure}[htbp]
    \centering
    \includegraphics[width=0.8\linewidth]{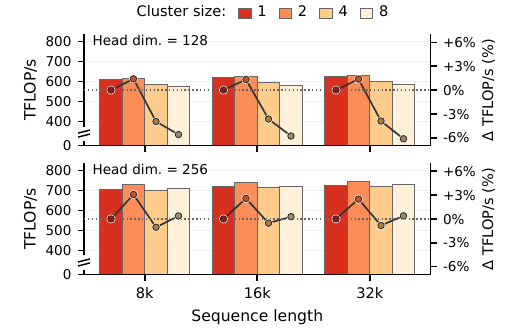}
    \caption{Increase in throughput of clustering over cluster size 1 across sequence lengths for Flash Attention.}
    \label{fig:fa4_seqlen_tflops}
\end{figure}

Figure \ref{fig:fa4_seqlen_tflops} shows that the TPC-level cluster size consistently achieves approximately 1.5\% and 3\% higher throughput than cluster size 1 for head dimensions of 128 and 256, respectively. However, as we see in Table \ref{tab:tensor_core_utilization}, the reduced data movement does not lead to increased tensor core utilization.
It is adjusted to account for the reduced active \gls{sm} limit ratio, as profiling results indicate a lower utilization at \gls{gpc}-level cluster sizes because it is averaged across all 132 \glspl{sm} in the \gls{gpu}. However, the active \gls{sm} limit restricts concurrent execution to 120 \glspl{sm}, leaving the remaining 12 \glspl{sm} idle and contributing 0\% utilization to the reported average. After accounting for these inactive \glspl{sm}, the tensor core utilization is comparable across cluster sizes. This indicates that the observed throughput improvements from reduced data movement are not driven by increased tensor core utilization.

\begin{table}[htb]
\caption{Tensor core utilization for Flash Attention. At cluster sizes 4 and 8, utilization is adjusted to the reduced \gls{sm} limit ratio.}
\label{tab:tensor_core_utilization}
\centering
\footnotesize
\renewcommand{\arraystretch}{0.8}
\begin{tabular}{llcccc}
\toprule
& & \multicolumn{4}{c}{Cluster size} \\
\cmidrule(lr){3-6}
Head dim. & Seq. len & 1 & 2 & 4 & 8 \\
\midrule
\multirow{2}{*}{128}
& 16K & 74.8\% & 74.3\% & 72.6\% & 69.5\% \\
& 32K & 76.8\% & 76.2\% & 75.3\% & 72.1\% \\
\midrule
\multirow{2}{*}{256}
& 16K & 94.3\% & 94.1\% & 92.5\% & 92.3\% \\
& 32K & 95.9\% & 95.8\% & 95.1\% & 95.2\% \\
\bottomrule
\end{tabular}
\end{table}

\begin{figure}[htbp]
    \centering
    \includegraphics[width=\linewidth]{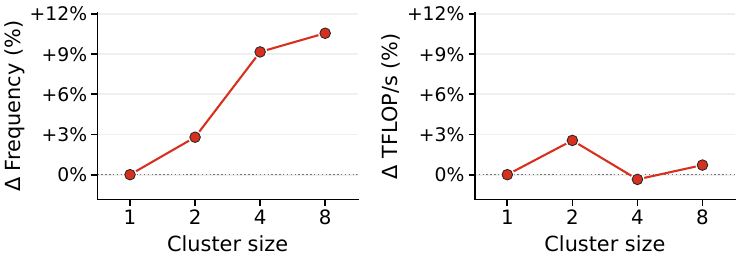}
    \caption{Percentage SM frequency and throughput improvement against cluster size 1 for FlashAttention 4 with 32K sequence length and 256 head dimension. Throughput increases proportionally to frequency variation at TPC-level clustering, and it does not follow in GPC-level clustering due to floorsweeping effects.}
    \label{fig:cluster_size_smfreq_hdim256_seqlen32k}
\end{figure}


We attribute these throughput improvements to the \gls{tdp} limit and the resulting frequency scaling through \gls{dvfs}. 
Figure \ref{fig:cluster_size_smfreq_hdim256_seqlen32k} shows that the \gls{sm} frequency increases with cluster size. This increase in frequency results in a proportional improvement in kernel throughput at the \gls{tpc}-level cluster size, i.e., cluster size 2. 
As illustrated in Fig. \ref{fig:tdp_dvfs_headroom}, increasing the cluster size reduces redundant data movement within the \gls{gpu}, consequently lowering the power consumed by data transfers. For kernels operating near the \gls{gpu}'s power limit, this reduction in power consumption creates additional power headroom, allowing the \gls{gpu}'s power management system to increase the \gls{sm} frequency through \gls{dvfs} while remaining within the \gls{tdp} limit.
However, this trend does not hold at \gls{gpc}-level cluster sizes, i.e., cluster sizes 4 and 8, because of the reduced active \gls{sm} limit discussed in Section~\ref{sec:nav_sm_limit_penalty}.

\subsection{Effect of TDP on other AI accelerators}
\vspace{-1mm}
To examine whether our observations generalize beyond GPUs, we further evaluate the impact of \gls{tdp} on Tenstorrent's Wormhole n300d~\cite{tenstorrent_wormhole_specs}, a representative spatial dataflow accelerator. 
The n300d card integrates two Wormhole chips, each with 64 Tensix cores at 1\,GHz, delivering a peak throughput of 64~TFLOP/s per chip.

\begin{figure}[htbp]
    \centering
    \includegraphics[width=\linewidth]{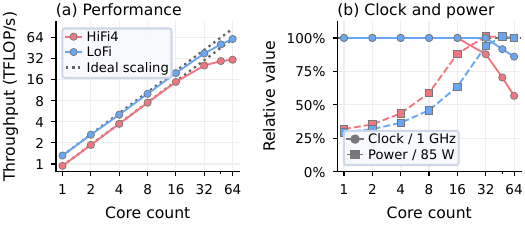}
    \caption{Scaling of matrix multiplication on one Wormhole chip under the
    default power limit. (a) Achieved throughput versus the number of active
    Tensix cores for the HiFi4 and LoFi math fidelities. (b) Average core clock
    (normalized to 1\,GHz) and power (normalized to the 85\,W limit).}
    \label{fig:tenstorrent_tdp}
\end{figure}

Figure~\ref{fig:tenstorrent_tdp} shows throughput, clock frequency, and power as the number of active Tensix cores grows, for two math fidelities: HiFi4, which makes multiple passes through the multiplier for full precision, and LoFi, which makes a single pass. With few active cores, both scale almost linearly at the nominal 1\,GHz clock. As more cores become active, power reaches the \gls{tdp} limit and the firmware lowers the clock to stay within it. HiFi4 uses more energy per operation and hits the limit first, at 32 cores. By 64 cores, its clock falls to 57\% of nominal, and its throughput levels off at about 31~TFLOPS, roughly half of ideal scaling. LoFi stays within the limit up to 32 cores and is throttled only to 86\% of nominal, reaching about 60~TFLOPS at 64 cores.

These results demonstrate that the impact of \gls{tdp} limits on kernel performance is non-trivial. Furthermore, understanding these effects provides greater insight into the observed performance characteristics. This reinforces our argument that microarchitectural knowledge alone is insufficient to maximize kernel performance or fully explain the performance achieved.
\section{Discussion}
\vspace{-1mm}
\subsection{Stringency and portability of wave-equalization}
\vspace{-1mm}
For a kernel launch to be wave-equalized across cluster sizes, it must satisfy Eq. \ref{eq:wave_equalized_kernel_launch_condition}. This requires a positive integer $n$ such that both $132(n-1) < B \le 132n$ and $120(n-1) < B \le 120n$ hold, where $B$ denotes the number of thread blocks and $n$ is the common number of execution waves. Combining these inequalities yields $132(n-1) < B \le 120n$, which is feasible only for $n < 11$. Therefore, the set of thread-block counts eligible for wave-equalized kernel launches is given by $\label{eq:wave_equalized_set}
 B\in\bigcup_{n=1}^{10}\bigl(132(n-1),\,120n\bigr]$.
This set is limited, but the limitation is unavoidable on the H100 because floorsweeping fixes the active \gls{sm} limits at 132 and 120. Considering kernel performance when making floorsweeping decisions could allow limits that expand this set. 
The condition also carries over to other \glspl{gpu}. The microbenchmark in Section~\ref{sec:active_sm_limits_at_diff_cluster_sizes} reports the number of unique \glspl{sm} observed per iteration. Unclustered runs (cluster size 1) give the complete active \gls{sm} limit, and \gls{gpc}-clustered runs give the reduced limit. Substituting these two values for 132 and 120 in Eq.~\ref{eq:wave_equalized_kernel_launch_condition} gives the condition for the target \gls{gpu}.

\subsection{Implications for kernel development systems}
\vspace{-1mm}
Compiler-based kernel optimization~\cite{chen2018tvm, zheng2020ansor, tillet2019triton}, manual performance engineering~\cite{dao2022flashattention, dao2023flashattention, shah2024flashattention}, and now agentic kernel development~\cite{ouyang2025kernelbench, han2026egg, ran2026kernelband, zhang2025cudaforge, dai2026cuda}, relies on appropriate abstractions and knowledge of the microarchitecture. The physical non-idealities discussed in this work challenge several assumptions underlying these approaches. For instance, widely accepted heuristics, such as amortizing memory accesses (e.g., amortizing L2 accesses through DSMEM) or increasing compute utilization up to saturation, can hurt performance. These effects also go beyond performance, potentially introducing numerical nondeterminism through changes in floating-point evaluation order, as well as power spikes. Through this work, we provide an initial impetus for the community to consider physical non-idealities more broadly. We derive a closed-form solution for manufacturing-yield effects on clustered execution and leave other non-idealities to future work.
\section{Conclusion}
\vspace{-1mm}
This work shows that physical non-idealities hidden by hardware abstractions can significantly affect kernel performance. On NVIDIA H100 \glspl{gpu}, floorsweeping restricts the number of \glspl{sm} available for clustered execution, increasing execution waves. Our wave-aware clustering strategy addresses these constraints, improving the throughput of the Hopper version of FlashAttention 4 by up to 5.30\% over the unclustered baseline. We also show that \gls{dsmem}-based reductions in data movement create power headroom for higher \gls{sm} frequencies, benefiting even compute-bound kernels under the \gls{tdp} limit. The influence of power constraints extends beyond \glspl{gpu}, as our analysis of Tenstorrent's Wormhole accelerator reveals similar limitations on throughput scaling. Together, these findings reinforce the need to consider physical non-idealities when analyzing and optimizing kernels across accelerator architectures.

\section*{Acknowledgments}

This work is partially supported by the Ministry of Education, Singapore, under Tier 3 grant MOET32024-0003 and the Advanced Research and Technology Innovation Centre (ARTIC), the National University of Singapore under Grant AFP-RP6.

\bibliography{main}
\bibliographystyle{mlsys2025}


\appendix


\end{document}